# Raman-induced Kerr-effect dual-comb spectroscopy

**Takuro Ideguchi [1], Birgitta Bernhardt [1,2], Guy Guelachvili [3], Theodor W. Hänsch [1,2] and Nathalie Picqué [1,2,3,*]**

1. Max-Planck-Institut für Quantenoptik, Hans-Kopfermann-Strasse 1, 85748 Garching, Germany
2. Ludwig-Maximilians-Universität München, Fakultät für Physik, Schellingstrasse 4/III, 80799 München, Germany.
3. Institut des Sciences Moléculaires d'Orsay, CNRS, Bâtiment 350, Université Paris-Sud, 91405 Orsay, France.
*Corresponding author e-mail address: nathalie.picque@mpq.mpg.de

*We report on the first demonstration of nonlinear dual-frequency-comb spectroscopy. In multi-heterodyne femtosecond Raman-induced Kerr-effect spectroscopy, the Raman gain resulting from the coherent excitation of molecular vibrations by a spectrally-narrow pump is imprinted onto the femtosecond laser frequency comb probe spectrum. The birefringence signal induced by the nonlinear interaction of these beams and the sample is heterodyned against a frequency comb local oscillator with a repetition frequency slightly different from that of the comb probe. Such time-domain interference provides multiplex access to the phase and amplitude Raman spectra over a broad spectral bandwidth within a short measurement time. Experimental demonstration, at a spectral resolution of 200 GHz, a measurement time of 293 µs and a sensitivity of $10^{-6}$, is given on liquid samples exhibiting a C-H stretch Raman shift.*

Optical frequency combs [1], initially developed for frequency metrology, are becoming enabling tools for a variety of applications in science and technology including broad spectral bandwidth molecular spectroscopy. New comb-based approaches to molecular spectroscopy are so far only probing the absorption and dispersion of the samples, mostly in the gas phase. Amongst these, dual-comb spectroscopy [2-8] holds much promise for rapid and sensitive precise recording of complex molecular spectra. The pulses of a frequency comb excite the sample at regular time intervals. The pulse train of a second comb of different repetition frequency samples the response of the medium, thus producing a time-domain interference pattern, whose Fourier transform yields the spectrum. Access to the fundamental vibrational bands of molecules is still one of the challenges to take up. Despite recent progress [2,5], mid-infrared dual-comb spectroscopy is impeded by a lack of convenient frequency comb generators [9]. Raman spectroscopy is an alternative technique to access fundamental molecular vibrations, which circumvents the need for mid-infrared lasers and photonics tools.

In this letter, we extend dual-comb spectroscopy to stimulated Raman scattering. To our knowledge, this is the first report of nonlinear dual-comb spectroscopy. Direct dual-comb emission spectroscopy and optically multi-heterodyne-detected femtosecond Raman-induced Kerr-effect spectroscopy (RIKE) are also demonstrated for the first time.

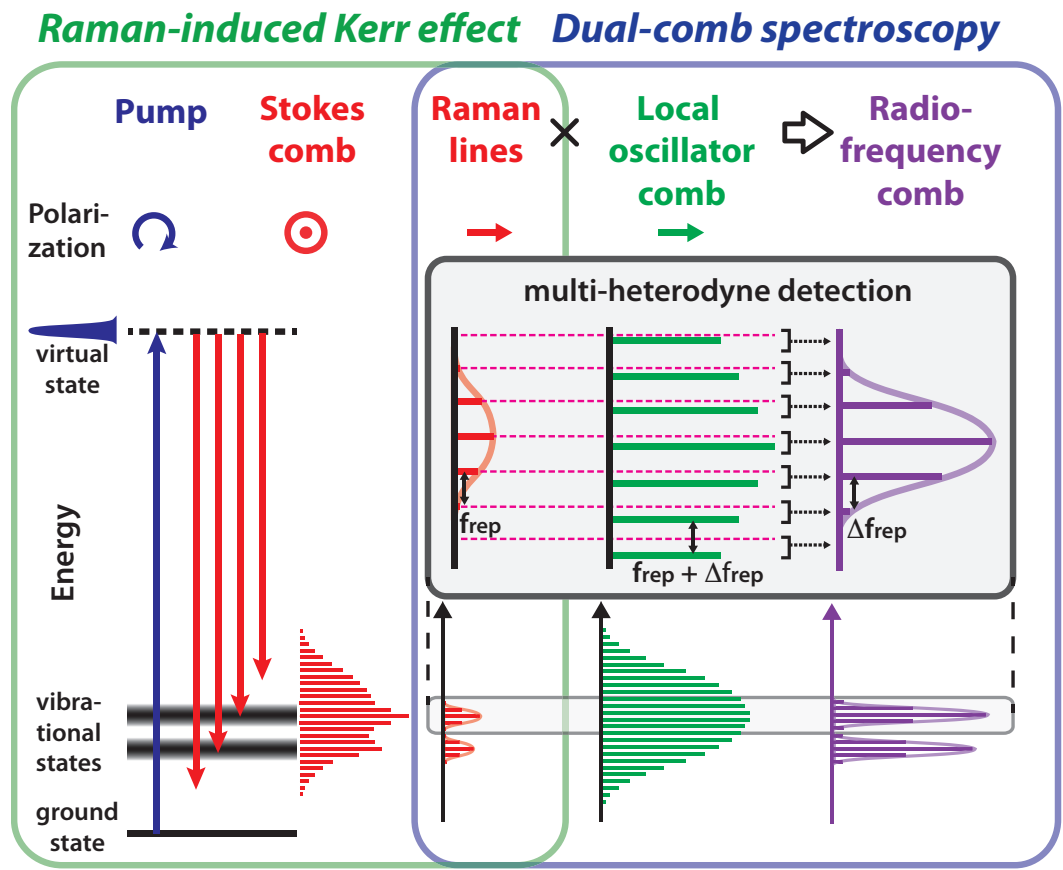

**Figure 1:** *Principle of Raman-induced Kerr-effect dual-comb spectroscopy.*

In RIKE dual-comb spectroscopy (Fig.1), a pump picosecond laser and a Stokes femtosecond frequency comb, with a repetition frequency of $f_{\text{rep}}$, are spatially and temporally overlapped onto the sample. Photons are transferred from the intense Raman pump laser to the weak Stokes beam when Stokes comb lines coincide with Raman resonances of the sample. The resulting frequency comb with the imprint of stimulated Raman transitions is heterodyned against a frequency comb similar to the Stokes comb, but with a slightly different repetition frequency, $f_{\text{rep}} + \Delta f_{\text{rep}}$. The multi-heterodyne detection magnifies the weak signal field and down-converts the comb of Raman signals to the radio-frequency domain, where it becomes easily accessible to fast digital processing. It provides simultaneous access to the amplitude and the phase of the stimulated Raman signals. This scheme without moving parts enables highly multiplexed fast measurements over a spectral span as broad as the bandwidth of the Stokes femtosecond optical comb. The instrumental resolution is only limited by the spectral width of the Raman pump laser. To enhance the sensitivity of our technique, we detect the changes in the polarization of the linearly-polarized Stokes comb induced at the sample by the circularly-polarized pump, in a scheme similar to femtosecond RIKE spectroscopy [10]. Such background-reduced detection also reduces artifacts like cross-phase modulation and parasitic interference patterns.

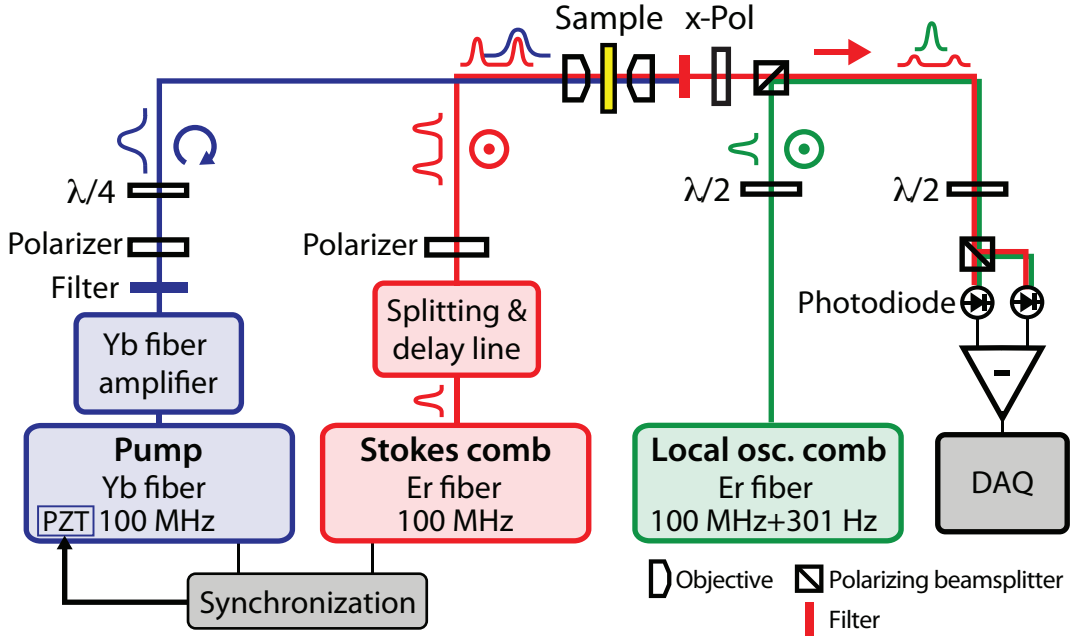

**Figure 2:** *Experimental setup for RIKE dual-comb spectroscopy. x-Pol: crossed-polarizer, λ/4: quarter-wave plate, λ/2: half-wave plate, PZT: piezo-electric transducer, DAQ: data acquisition.*

The experimental setup is displayed in Fig. 2. A femtosecond Yb-fiber based laser/amplifier system [4] with a repetition frequency of 100 MHz serves as the Raman pump. Its emission is centered at a wavelength of 1.04 μm, with a full-width at half-maximum (FWHM) of 37 cm$^{-1}$ after spectral filtering, and the pulses are slightly chirped. The pump beam is circularly polarized. The Stokes probe laser is a linearly polarized Er-fiber frequency comb emitting about 20 mW of average power and centered at 1.55 μm with a FWHM of 340 cm$^{-1}$ that limits the spectral range in this proof-of-principle experiment but enables stimulation of Raman signals with shifts around 3000 cm$^{-1}$, i.e. in the C-H stretch region. The repetition frequency of the Yb pump laser is actively synchronized to the Er comb using a scheme similar to the one reported in [11]. The fundamental repetition frequency and its 100th harmonic are compared to those of the Er-fiber laser. The two resulting error signals are combined through a continuously variable potentiometer and fed back to a piezo-electric transducer inside the Yb oscillator cavity. Sum frequency generation cross-correlation measurements of the spectrally-filtered Yb beam and the Er oscillator output, performed over a time period of a few seconds using two different electronic filtering bandwidths (1 MHz and 160 Hz), provide a root-mean-square timing jitter of 34 and 15 fs, respectively. The output of the erbium oscillator is split to generate a pulse that is temporally overlapped with the Raman pump and a reference pulse used to measure the spectrum of the Stokes laser without pump. The two pulses are separated by about 250 ps. An infinity-corrected objective focuses the pump (diameter: 6 μm, Rayleigh length: 28 μm) and Stokes beams (diameter: 7 μm, Rayleigh length: 22 μm) into a 1 mm thick cuvette containing the liquid sample. The positions of the focal points of the pump and Stokes beams are located about 40 μm apart due to chromatic aberrations of the microscope objective, and this reduces the efficiency of the stimulated Raman process. At the sample, the pump and Stokes beams have an average power of 960 mW (energy of 9.6 nJ) and 1.9 mW (energy of 9.5 pJ), respectively. After recollimation of the beams, the pump light is rejected by a long-wavelength-pass optical filter. A cross-polarizer is aligned to block the Stokes beam in the absence of a pump. The optical birefringence induced in the initially isotropic medium by the anisotropic Raman resonant third-order polarization enables transmission of the background-reduced comb of Raman resonances. The extinction ratio of the Stokes probe through the entire optical set-up is $10^{-4}$, mostly limited by the microscope objectives and the sample cell. A second femtosecond Er-fiber comb, with a repetition frequency detuned by 301 Hz and an average output power of 20 mW, optically samples the trains of pulses transmitted by the polarizer. The amplitude of this local oscillator for multi-heterodyne detection is controlled by the combination of a half-wave plate and a polarizing beam-splitter to optimize the signal to noise ratio in the interferogram. The time-domain interference signal is monitored with a differential InGaAs photodetector, which collects the two outputs of the interferometer. After electronic filtering, the interferometric pattern only contains the Raman gain signal, allowing for efficient amplification. The dynamic range issues associated with dual-comb spectroscopy are therefore partially overcome. Two interferograms, time-delayed by 83 µs, are sequentially sampled at 100 MSamples/s by a 14-bit data acquisition board. Their complex Fourier transform reveals the phase and amplitude spectra in presence and absence of the pump laser. The ratio of the spectra with and without pump simultaneously provides the Raman gain spectrum of the dispersive (real part) and resonant (imaginary part) tensor elements of the third-order nonlinear susceptibility.

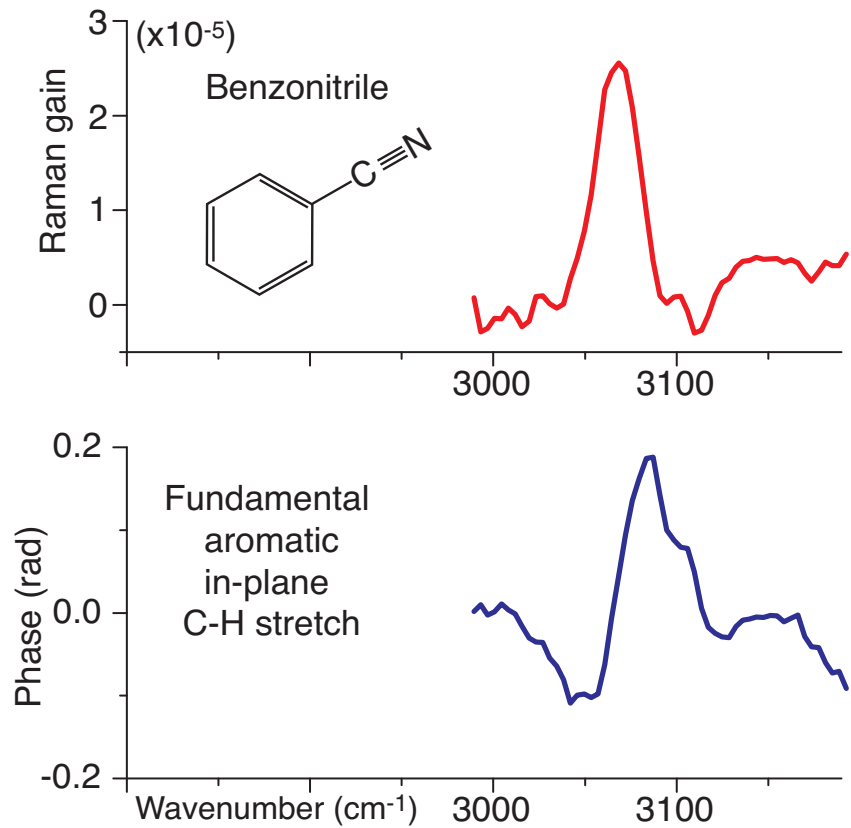

*Figure 3: Experimental Raman gain amplitude and dispersion spectra of benzonitrile, measured within 293 µs.*

An interferogram with 200 GHz resolution is measured within 2.93 µs and the acquisition rate of the interferograms is the difference in repetition frequency of the two erbium lasers, 301 Hz, chosen according to the free spectral range span and the comb line spacing. To get a reasonable signal-to-noise ratio, several interferograms are averaged. Figure 3 displays a portion of the amplitude and phase spectra of neat benzonitrile, featuring the Raman band of the in-plane C-H stretch centered at 3073 cm$^{-1}$, at an instrumental resolution of 6.7 cm$^{-1}$. To compute these spectra, one hundred interferograms are averaged, resulting in an effective measurement time of 293 µs and a total experimental time of 332 ms. The observed linewidth (about 27.5 cm$^{-1}$) of the band is narrower than the width of the Raman pump laser as the pulses of the latter are chirped. The stimulated Raman gain experienced by the Stokes beam at the sample is 10$^{-5}$. The relative noise level in our detection scheme is 10$^{-6}$. The weaker bands of p-xylene corresponding to the methyl and aromatic C-H stretch are recorded with the same experimental set-up. One thousand and sixty-two interferograms, each measured within 2.99 µs, are averaged, leading to an effective measurement time of 3.2 ms and an experimental time of 3.5 s. The noise level in our experiment already represents an excellent figure for real-time femtosecond stimulated Raman spectroscopy: the improvement over the experiment with a high-repetition frequency system reported in [12] is of two orders of magnitude, while our measurement time is 5 times shorter. Sensitivities better than 10$^{-8}$ have been reported [13] with lock-in detection schemes, but these only allow for multiplex detection of three spectral elements. Most femtosecond stimulated Raman experiments (e.g. [10]) use complex amplified laser systems with kHz-Hz repetition frequencies and long measurement times. Our low-intensity set-up and our very short measurement time therefore open intriguing opportunities for real-time spectrally resolved Raman chemical labeling. The sensitivity of our experimental scheme could be significantly improved by achieving a better spatial overlap between the pump and Stokes beams, either by the implementation of reflective microscope objectives or by use of laser systems, like Ti:Sa oscillators, that emit in spectral regions where microscopy instrumentation is more advanced.

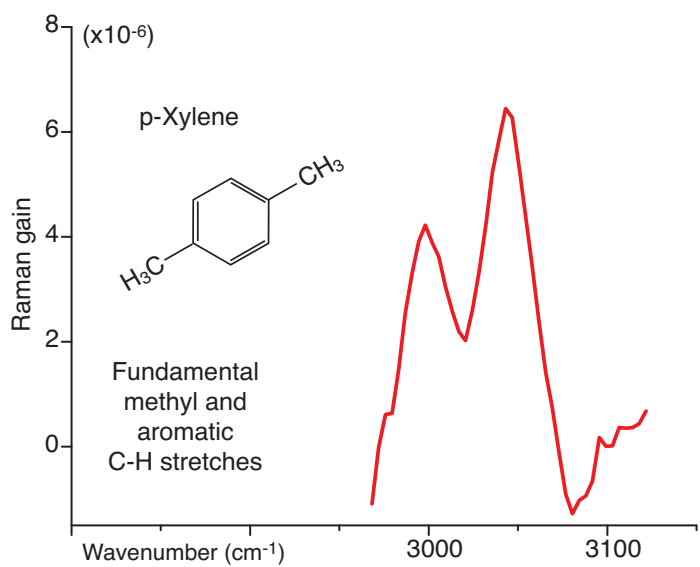

***Figure 4:*** *Experimental spectrum of p-xylene showing the resolved bands corresponding to the methyl and aromatic C-H stretch.*

Much faster acquisition rates may be obtained if the difference in repetition frequency of the two combs is varied during the experiment [8] or if the line spacing of the combs matches the desired spectral resolution. Frequency comb generators produced by four-wave mixing in microresonators [14] typically exhibit a line spacing ranging between 10 and 500 GHz, and might therefore open up a route to on-chip spectrometers for liquid-phase Raman spectroscopy. As the interaction length for the stimulation of Raman transitions is a few tens of micrometers, our instrument suits very well the requirements of sensitive microscopic mapping or imaging of chemical or biological samples, as well as capillary electrophoresis and microfluidic devices.

Support by the Max Planck Foundation, the European Laboratory for Frequency Comb Spectroscopy, the Munich Center for Advanced Photonics, the European Research Council (Advanced Grant 267854), Eurostars and the Agence Nationale de la Recherche are acknowledged.

**References:**
[1] T.W. Hänsch, Nobel Lecture: Passion for Precision, Rev. Mod. Phys. **78**, 1297-1309 (2006).
[2] E. Baumann, F.R. Giorgetta, W.C. Swann, A.M. Zolot, I. Coddington, N.R. Newbury, Spectroscopy of the methane $\nu_3$ band with an accurate midinfrared coherent dual-comb spectrometer, Phys. Rev. A **84**, 062513 (2011).
[3] T. Ideguchi, A. Poisson, G. Guelachvili, N. Picqué, T.W. Hänsch, Adaptive real-time dual-comb spectroscopy, arXiv:1201.4177 (2012).
[4] B. Bernhardt, A. Ozawa, P. Jacquet, M. Jacquey, Y. Kobayashi, T. Udem, R. Holzwarth, G. Guelachvili, T.W. Hänsch, N. Picqué, Cavity-enhanced dual-comb spectroscopy, Nat. Photon. **4**, 55-57 (2010).
[5] B. Bernhardt, E. Sorokin, P. Jacquet, R. Thon, T. Becker, I. T. Sorokina, N. Picqué, T.W. Hänsch, Mid-infrared dual-comb spectroscopy with 2.4 $\mu$m $Cr^{2+}$:ZnSe femtosecond lasers, Appl. Phys. B **100**, 3-10 (2010).
[6] J.D. Deschênes, P. Giaccarri, J. Genest, Optical referencing technique with CW lasers as intermediate oscillators for continuous full delay range frequency comb interferometry, Opt. Express **18**, 23358 (2010).
[7] I. Coddington, W.C. Swann, N.R. Newbury, Coherent multiheterodyne spectroscopy using stabilized optical frequency combs, Phys. Rev. Lett. **100**, 013902 (2008).
[8] A. Schliesser, M. Brehm, F. Keilmann, D.W. van der Weide, Frequency-comb infrared spectrometer for rapid, remote chemical sensing, Opt. Express **13**, 9029-9038 (2005).
[9] A. Schliesser, N. Picqué, T.W. Hänsch, Mid-infrared frequency combs, Nat. Photon. **6**, 440-449 (2012).
[10] S. Shim, R.A. Mathies, Femtosecond Raman-induced Kerr effect spectroscopy, J. Raman Spectrosc. **39**, 1526–1530 (2008).
[11] D.J. Jones, E.O. Potma, J. Cheng, B. Burfeindt, Y. Pang, J. Ye, X.S. Xie, Synchronuzation of two-passively modelocked, picosecond lasers within 20 fs for

coherent anti-Stokes Raman scattering microscopy, Rev. Sci. Instrum. **73**, 2843-2848 (2002).
[12] E. Ploetz, B. Marx, T. Klein, R. Huber, P. Gilch, A 75 MHz Light Source for Femtosecond Stimulated Raman Microscopy, Opt. Express **17**, 18612-18620 (2009).
[13] C.W. Freudiger, W. Min, G.R. Holtom, B. Xu, M. Dantus, X.S. Xie, Highly specific label-free molecular imaging with spectrally tailored excitation-stimulated Raman scattering (STE-SRS) microscopy, Nat. Photon. **5**, 103-109 (2011).
[14] C. Wang, T. Herr, P. Del'Haye, A. Schliesser, R. Holzwarth, T. W. Hänsch, N. Picqué, T.J. Kippenberg, Mid-Infrared Frequency Combs Based on Crystalline Microresonators, arXiv:1109.2716 (2011).